\documentclass[aps,prl,reprint,floatfix,citeautoscript,longbibliography,superscriptaddress]{revtex4-1}

\usepackage[english]{babel}
\RequirePackage[T1]{fontenc}
\RequirePackage{times} 

\usepackage{siunitx} 
\usepackage[italicdiff]{physics}
\usepackage{amsfonts}
\usepackage{subfigure}
\usepackage{amsmath}
\usepackage{amssymb}
\usepackage{graphicx,epstopdf}
\usepackage{xspace}
\usepackage{array}
\usepackage[hidelinks]{hyperref}
\usepackage{xcolor}
\usepackage{my_symbols}
\usepackage{CJK}
\usepackage{bm}
\usepackage{verbatim}
\usepackage{lipsum}
\usepackage{tikz}
\usepackage[normalem]{ulem}

{\par}

\providecommand{\U}[1]{\protect\rule{.1in}{.1in}}

\newcounter{mycomment}

\newcommand\rmv{\bgroup\markoverwith {\textcolor{red}{\rule[0.5ex]{2pt}{0.4pt}}}\ULon}

\begin{document}
\begin{CJK*}{UTF8}{gbsn}
\title{A loop theory for the input-output problems in cavities}
\author{H. Y. Yuan}
\email[Corresponding author:~]{huaiyangyuan@gmail.com}
\affiliation{Department of Physics, Southern University of Science and Technology, Shenzhen 518055, China}
\author{Weichao Yu (余伟超)}
\affiliation{Institute for Materials Research, Tohoku University, Sendai 980-8577, Japan}
\affiliation{Department of Physics and State Key Laboratory of Surface Physics, Fudan University, Shanghai 200433, China}
\author{Jiang Xiao (萧江)}
\email[Corresponding author:~]{xiaojiang@fudan.edu.cn}
\affiliation{Department of Physics and State Key Laboratory of Surface Physics, Fudan University, Shanghai 200433, China}
\affiliation{Institute for Nanoelectronics Devices and Quantum Computing, Fudan University, Shanghai 200433, China}
\date{\today}

\begin{abstract}
The input-output formalism is the basis to study the response of an optical cavity to the external stimulations. The existing theories usually handle cavity systems with only one internal mode. However, there is growing interest in more complex systems, especially the hybrid cavity-matter systems, which contains at least two internal modes, one or more from the optical cavity and the matter, respectively.
Here we propose a graphical loop theory to calculate and visualize the reflection and transmission spectrum of such multi-mode cavity, resembling the role of Feynman diagrams in the quantum field theory. This loop theory gives a unified picture to interpret the experimental observations on a hybrid magnet-light system, and is extremely easy to apply to arbitrary complicated problems without any calculations.
\end{abstract}

\maketitle
\end{CJK*}

{\it Introduction.} Exploring the nature of light and its interaction with condensed matters is a long-lasting topic in both optical and condensed matter physics. An important knob to manipulate light
is to guide it into mediums with finite volume such as cavity, waveguide or other types
of resonators. This route has inspired fruitful physical concepts, including the light
quantum (photon), the squeezed state, quantum information, as well as numerous applications
such as laser, optical fiber, and optical tweezer. To probe the light-matter interaction,
the typical practice is to put the matter in a cavity, and analyze the response
(the transmission and reflection probabilities) of the light modulated by the hybrid
matter-cavity system. This idea has given birth to the known fields such as cavity
quantum electrodynamics \cite{Walther2006}, cavity opto-mechanics \cite{Asp2014}, and
cavity spintronics or spin cavitronics \cite{Dany2019}. To interpret the reflection
and transmission spectrum of light in these hybrid systems, it is essential to have
a theoretical formalism to connect the input and output electromagnetic waves with
the internal cavity modes. For a single-mode cavity, this relation has been well
established \cite{Collett1984,DFWalls,Garrison},
where the input wave stimulates the cavity mode via a Hermitian interaction,
which in turn also serves as a generic decay for the cavity mode. Such a theory
is sufficient to analyze the Lorentz transmission spectrum of a cavity when there
is only one cavity mode is relevant, while the other modes are either largely
detuned or not directly excited by the input wave. The situation becomes more
complicated for a hybrid cavity-matter system, where the hybridized modes are
close in energy and may be directly excited simultaneously. Take the cavity-magnet
system as an example, the dipolar fields produced by the precessing magnetization
inevitably mix with the cavity standing wave, and it is the superposition wave that
couples with the probe light \cite{Auld1963,Soykal2010,Cao2015,Yu2019}.
In such situations, the traditional input-output theory is insufficient, especially
when the input wave couples directly with more than one internal mode.

In this Letter, based on the basic principles of quantum mechanics, we propose a
universal loop theory to analyze the reflection and transmission spectrum of a multi-mode
cavity. An analytical expression of the transmission, as a function of mode frequency,
coupling coefficients among the cavity modes, and dissipation strength of cavity modes,
are explicitly derived.
The loop theory provides an extremely simple graphical approach to solve the input-output problems in all types of cavity-based systems without carrying out tedious calculation. Because of its simplicity, the loop theory also presents a unified physical picture to understand the experimental results in different regimes, such as the asymmetric Fano resonance, Purcell effect, repulsive anti-crossing in strong coupling regime, as well as the attractive level crossing in dissipative coupling regime.



\begin{figure}[b]
  \centering
 \includegraphics[width=\columnwidth]{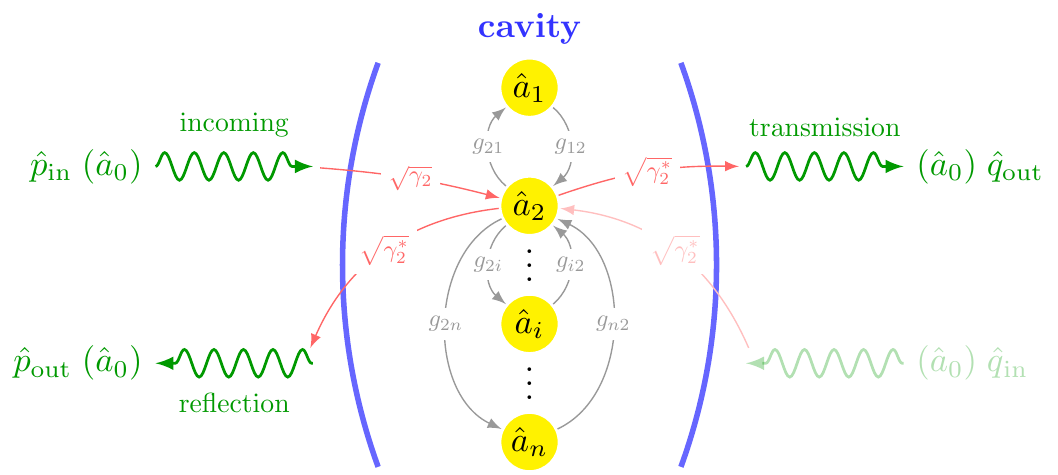}
  \caption{A multi-mode cavity (modes as $\ha_{1,2,\dots, n}$) is connected with input and output via the $\ha_0$ mode. The cavity modes are mutually coupled with one another via coupling strength $g_{ij}$, and coupled with the input (output) via $\sqrt{\gamma_i}$ ($\sqrt{\gamma_i^*}$). This figure only shows the coupling connections to the $\ha_2$ mode. }
  \label{fig:cavity}
\end{figure}

{\it Physical model.}
We consider an $n$-mode cavity, as shown in Fig. \ref{fig:cavity}, where the internal mode represented by operators $\ha_i ~(i=1,2,...n)$ has natural frequencies $\omega_i = \nu_i + i \eta_i$ with $\nu_i$ and $\eta_i$ being real and imaginary (dissipative) parts. These modes are not necessarily the optical modes of the cavity, but can also be the modes from the matter placed in and/or coupled with the cavity. The cavity is connected to external ports, which guide the external wave into/out of the cavity, and the external incoming/outgoing wave is represented by operator $\ha_0$, whose frequency $\omega$ can be scanned. The coupling strength between two cavity modes (mode-$i$ and -$j$) is $g_{ij}$, while the coupling strength between mode-$i$ and the external mode $\ha_0$ is $\sqrt{\gamma_i}$. All the coupling coefficients are presumably complex.
Depending on the amplitude of coupling strength with the external mode $\sqrt{\gamma_i}$, cavities can be classified into two categories, \ie the open cavities with large openings, such as the Fabry-Perot cavity and the cross-line cavity \cite{Harder2018,Bohi2019}, and the closed cavities with small openings, such as the CPW resonator, reentrant cavity and rectangular cavity \cite{Huebl2013,Gor2014,Bai2015}.

The Hamiltonian of the multi-mode cavity system includes the contributions from the internal cavity modes ($\hH_c$), the external mode ($\hH_e$), and their interactions ($\hH_\mathrm{int}$):
\begin{equation}
\hH= \hH_{c} + \hH_e + \hH_{\mathrm{int}}.
\label{ham}
\end{equation}
Here $\hH_{c}$ is a functional of the internal boson modes inside a cavity and their mutual interaction, \ie
\begin{equation}
\hH_{c} = \sum_{i=1}^n \hbar\omega_i \ha^\dagger_i \ha_i
+ \frac{\hbar}{2}\sum_{i\neq j}  \qty(g_{ij}\ha^\dagger_i \ha_j + g_{ij}^*\ha_i \ha^\dagger_j),
\end{equation}
where $\ha_i^\dagger$ ($\ha_i$) are the creation (annihilation) operators of the $i$-th cavity mode, satisfying boson commutation relations $[\ha_i,\ha_j^\dagger]=\delta_{ij}$, and $g_{ji}=g_{ij}^*$ is imposed to guarantee the Hermitian nature of the system. $\hH_e$ is the free Hamiltonian of the external reservoir fields,
\begin{equation}
\hH_e
= \int \dd{\omega} \hbar\omega  \ha^\dagger_0 (\omega) \ha_0 (\omega),
\end{equation}
where $\ha_0^\dagger(\omega)$ ($\ha_0(\omega)$) are the creation (annihilation) operator of the external
field of frequency $\omega$, satisfying the commutation relation
$[\ha_0(\omega),\ha_0^\dagger(\omega')]=\delta(\omega-\omega')$.
$\hH_{\mathrm{int}}$ is the interaction between the cavity modes
and the external fields,
\begin{equation}
\hH_{\mathrm{int}}
= \frac{\hbar}{\sqrt{2\pi}}\sum_i \int \dd{\omega} \qty[\sqrt{\gamma_i^*(\omega)}~\ha^\dagger_i \ha_0(\omega) + \mbox{c.c.}],
\label{couple}
\end{equation}
where $\sqrt{\gamma_i(\omega)}$ is the coupling amplitude between the $i$-th cavity mode and the
external field at frequency $\omega$.

\begin{figure}[t]
  \centering
 \includegraphics[width=0.9\columnwidth]{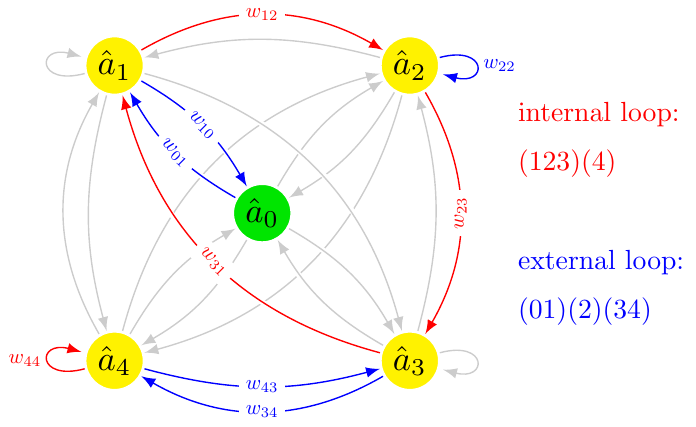}
  \caption{The graph representation of a four-mode cavity ($n=4$). The yellow vertices represent the cavity modes while the green vertex represents the input/output mode. The edges connecting the vertices represent the effective coupling between the modes. The edges highlighted in red and blue represents an internal and an external loop, denoted as $(123)(4)$ and $(01)(2)(34)$.}
  \label{fig:graph4}
\end{figure}

{\it The graph representation.} To state our principle result, we first depict the multi-mode system as a complete {\it weighted directed graph} $G_n$ with $n+1$ vertices. An example of such graph for $n=4$ is shown in \Figure{fig:graph4}. The vertices of the graph $V(G)$ represent all the physical internal/external modes, and $E(G)$ includes all directed edges $e_{ij}$ from $\ha_i$ to $\ha_j$:
\begin{subequations}
\begin{align}
V(G_n) &= \qty{\ha_0,\ha_1,\ha_2,\dots,\ha_n}, \\
E(G_n) &= \qty{e_{ij}~|~i,j=0,\dots n}.
\end{align}
\end{subequations}
The weight of the directed edge represents the effective coupling strength between the connected modes:
\begin{equation}
W(e_{ij}) =  w_{ij} = \kappa_{ij}\sqrt{\Delta_i\Delta_j}.
\end{equation}
Here $\kappa_{ij}$ is the renormalized coupling strength defined as $\kappa_{ij} = g_{ij} - i\sqrt{\gamma_i\gamma_j^*}$ for $i\neq j \neq 0$ (note $\kappa_{ij}\neq \kappa_{ji}^*$ in general), $\kappa_{0i} = \kappa_{i0}^* = \sqrt{\gamma_i}$, $\kappa_{00} = 0$, and $\kappa_{ii} = \Delta_i^{-1} \equiv \omega_i-\omega-i\gamma_i$ represents the detuning of the cavity mode from the external input/output mode, and $\Delta_0 \equiv 1$.

We define an $m$-loop $F_m$ as a subgraph of $G_n$ with $m$ vertices, and each vertex has exactly one incoming edge and one outgoing edge. Such an $m$-loop is in fact a union of disjoint cyclic paths connecting $m$ vertices, therefore we can use a cyclic decomposition of a permutation to represent an $m$-loop. For instance, for the case of $n=4$ as depicted in \Figure{fig:graph4}, a $4$-loop, consisting of two disjoint cycles of $\ha_1\ra\ha_2\ra\ha_3\ra\ha_1$ and $\ha_4\ra\ha_4$, can be represented as a permutation cycle $F_4 = (123)(4)$ or equivalently as $(231)(4)$ or $(312)(4)$. A $5$-loop, formed by three disjoint cycles of $\ha_1\ra\ha_0\ra\ha_1$, $\ha_2\ra\ha_2$, and $\ha_3\ra\ha_4\ra\ha_3$, is represented as $F_5 = (01)(2)(34)$.
For the purpose of this Letter, we are only interested in two particular types of loops: (i) the internal $n$-loop, $F_n$, which involves all $n$ cavity modes but not the external mode, \ie $V(F_n) = \qty{\ha_1,\dots, \ha_n}$; (ii) the external $(n+1)$-loop, $F_{n+1}$, which involves all $n$ cavity modes and the external mode, \ie $V(F_{n+1}) = \qty{\ha_0,\ha_1,\dots, \ha_n}$. In general, there are $n!$ internal loops and $n\cdot n!$ external loops. The $4$-loop $(123)(4)$ and the $5$-loop $(01)(2)(34)$ in \Figure{fig:graph4} are the examples for the internal and external loops for $n=4$.

\begin{figure*}[t]
  \centering
  \includegraphics[width=0.98\textwidth]{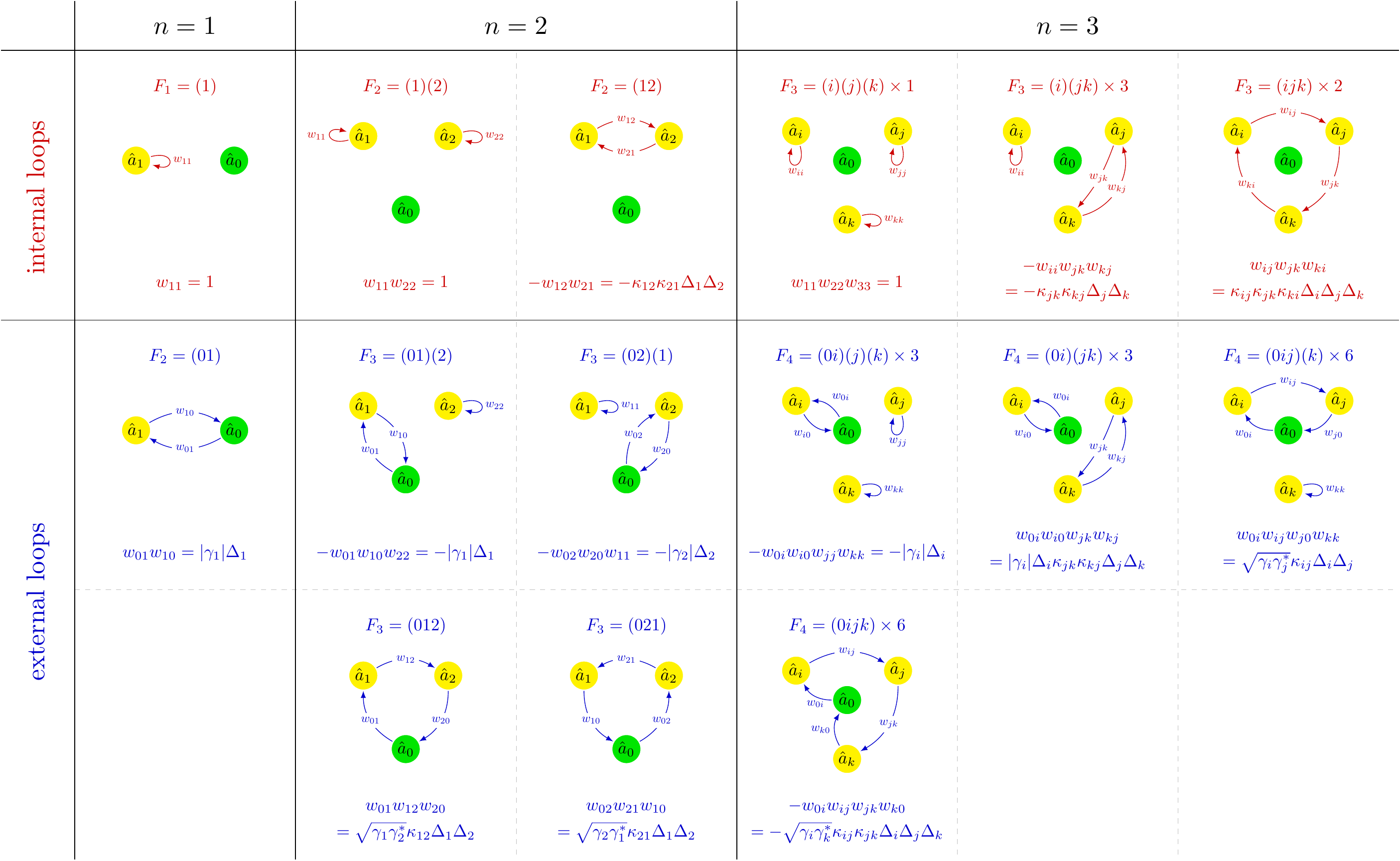}
  \caption{All internal (in red) and external (in blue) loops for the case of $n= 1, 2, 3$. The $\cA$ value for each loop is calculated using \Eq{eqn:AF}. The number following the $\times$ symbol for the $n=3$ case indicates the number of inequivalent patterns due to different permutation of $(i,j,k)$. }
  \label{fig:loops}
\end{figure*}

{\it The loop theorem.} Each $m$-loop $F_m$ is assigned with a dimensionless amplitude $\cA(F_m)$ given by the product of the weight of all edges of the loop:
\begin{equation}
  \label{eqn:AF}
  \cA(F_m) = P(F_m)\prod_{e_{ij}\in E(F_m)} W(e_{ij}).
\end{equation}
Here $P(F_m) = (-1)^{m-k}$ is the parity of permutation $F_m$, where $k$ is the number of disjoint cycles (brackets) of $F_m$.

The transmission spectrum for an open cavity system represented by graph $G$ can be simply expressed as:
\begin{equation}
\label{theorem}
S_{21}(G_n) = 1 + i~\frac{-\displaystyle \sum_\ssf{external} \cA(F_{n+1})}{\displaystyle\sum_\ssf{internal} \cA(F_n)}.
\end{equation}
The transmission for a closed cavity is given by the same expression but with the unity removed. The detailed proof of the theorem can be found in the Supplemental Materials \cite{notesp}. The loop theorem in \Eq{theorem} is the central result of this Letter.

Physically, the sum of amplitude of the external loops corresponds to the superposition of different propagation channels, similar to the sum of all Feynman diagrams in a scattering process in quantum field theory. While the internal loops in the denominator represent the renormalization from the internal scattering process.

{\it Application of the loop theorem.} Let us apply the loop theorem to study the transmission spectrum for three examples with $n=1,2,3$, respectively. Based on the same rules, it is straightforward to apply the loop theorem to find the transmission for arbitrary $n$.

For a single-mode cavity ($n=1$), there is only one internal (self-)loop and one external loop as shown in \Figure{fig:loops}. Plugging the corresponding $\cA$ value for each loop from \Eq{eqn:AF} in the loop theorem \Eq{theorem}, one obtains
\begin{equation}
  \label{eqn:S21n1}
S_{21}(G_1) = 1+i \abs{\gamma_1}\Delta_1 =1+i\frac{|\gamma_1|}{\omega_1-\omega- i\gamma_1}.
\end{equation}
When $\gamma_1$ is real, \Eq{eqn:S21n1} reduces to the well-known Lorentz type transmission spectrum for a single-mode cavity \cite{DFWalls} with a total linewidth of $\eta'_1 = \eta_1 + \gamma_1$, representing the intrinsic damping $\eta_1$ and the extrinsic damping due to leakage through the cavity ports.

For a two-mode cavity ($n=2$), there are two internal and four external loops (see \Figure{fig:loops}). Plugging the $\cA$ value for each loop in the loop theorem \Eq{theorem}, one obtains
\begin{align}
\label{eqn:S21n2}
S_{21}(G_2) =1 &+ i\frac{|\gamma_1|\Delta_1 + |\gamma_2| \Delta_2 }{1-\kappa_{12}\kappa_{21}\Delta_1\Delta_2} \\
&+i\frac{- \qty(\kappa_{12}\sqrt{\gamma_2^*\gamma_1}+\sqrt{\gamma_1^*\gamma_2}\kappa_{21})\Delta_1\Delta_2}{1-\kappa_{12}\kappa_{21}\Delta_1\Delta_2}. \nonumber
\end{align}
This expression has the parity symmetry ($1 \leftrightarrow 2$). In the most widely studied scenarios where only one out of the two internal modes is coupled to the external field, \eg $\gamma_2 > 0$ and $\gamma_1=0$, \Eq{eqn:S21n2} reduces to
\begin{equation}
\label{eqn:S21n2b}
S_{21}(G_2) = 1-\frac{i\gamma_2}{(\omega-\omega_2+i\gamma_2)-g^2_{12}/(\omega-\omega_1)},
\end{equation}
which is the widely used formula possessing the typical avoided level anti-crossing feature for the two modes \cite{Cao2015}.

For a three-mode cavity ($n=3$), there are three types (6 in total) of internal loops and four types (18 in total) of external loops (see \Figure{fig:loops}). By summing up all the contributions according to the loop theorem \Eq{theorem}, one obtains
\begin{widetext}
\begin{equation}
\label{eqn:S21n3}
S_{21}(G_3) = 1+i\frac{|\gamma_1|\Delta_1(1-\kappa_{23}\kappa_{32} \Delta_2 \Delta_3)+|\gamma_2|\Delta_2(1-\kappa_{13}\kappa_{31} \Delta_1 \Delta_3)
+|\gamma_3|\Delta_3(1-\kappa_{12}\kappa_{21} \Delta_1 \Delta_2) + I}{1-\kappa_{12}\kappa_{21} \Delta_1 \Delta_2 - \kappa_{13}\kappa_{31} \Delta_1 \Delta_0 - \kappa_{23}\kappa_{32} \Delta_2 \Delta_3 + (\kappa_{13}\kappa_{21} \kappa_{32}+\kappa_{12}\kappa_{23}\kappa_{31}) \Delta_1 \Delta_2 \Delta_3}.
\end{equation}
\end{widetext}
Here
\begin{align*}
I\equiv &-\frac{1}{2}\sum_{i\neq j}\qty(\sqrt{\gamma_i^* \gamma_j}\kappa_{ji} + \kappa_{ij}\sqrt{\gamma_j^*\gamma_i} ) \Delta_i \Delta_j \\
&+ \frac{1}{2}\sum_{i\neq j\neq k}\qty(\kappa_{ki}\sqrt{\gamma_i^* \gamma_j}\kappa_{jk}  +\kappa_{kj}\sqrt{\gamma_j^*\gamma_i} \kappa_{ik}) \Delta_i\Delta_j\Delta_k
\end{align*}
represents the interference effect of different scattering paths when two
or more internal modes are coupled with the external wave simultaneously.



{\it Physical implications.} For a two-mode cavity case, when only one cavity mode ($\gamma_2 \neq 0, \gamma_1 = 0$) is driven by the input, the transmission can be well described by \Eq{eqn:S21n2b}. Depending on the relative strength of the coupling strength ($g_{12}$) and the total dissipation of each cavity mode ($\eta'_1,\eta'_2$), four types of regimes can be identified (see Supplementary Materials for a detailed discussion):
i) the strong coupling regime with a typical avoided level crossings for $g_{12} > \eta'_1, \eta'_2$, ii) asymmetric Fano lineshape with electromagnetically induced transparency (EIT/MIT) at resonance for $\eta'_2 > g_{12} > \eta'_1$, iii) the Purcell regime for $\eta'_2 < g_{12} < \eta'_1$, and iv) the weak coupling regime for $g_{12} < \eta'_1, \eta'_2$, all of which have been demonstrated in experiments \cite{Limo2017,Zhang2014,note4phase}. In fact, the seemingly simple two-mode cavity system contains much rich physics than the well-studied four regimes above. An apparent unexplored realm would be that both cavity modes are coupled to the external fields, \ie $\gamma_1, \gamma_2 \neq 0$, then the interference terms (second line in \Eq{eqn:S21n2}) shall manifest itself.
Even more versatile variants would be considering the relative phase of $\gamma_1, \gamma_2$. All of these features are contained in the complete transmission expression \Eq{eqn:S21n2} for two-mode cavity obtained by the loop theorem, and yet to be explored.

An application of the three-mode cavity involves the attractive level crossing between magnon and photon modes observed in an opto-magnetic cavity system by Harder \etal \cite{Harder2018}. To explain the non-conventional attractive level crossing, Yu \etal proposed a minimal three-mode model including a magnon mode ($\ha_1$), a cavity mode ($\ha_2$) of low dissipation, and another hidden cavity mode ($\ha_3$) of strong dissipation \cite{Yu2019}. The transmission spectrum of such three-mode cavity can be well captured by \Eq{eqn:S21n3}.
The strong dissipation of the mode $\ha_3$ simplifies the story by smearing out the interference effect, which is equivalent to neglect the interference term $I \ra 0$. Consequently, as demonstrated in Ref. \onlinecite{Yu2019}, \Eq{eqn:S21n3} can reproduce both repulsive and attractive level crossing behaviors observed in the experiments in such systems. The core physics is that the high dissipation mode ($\ha_3$), functioning as an effective delay line for the coupling, can mediate both real and imaginary couplings between the cavity mode ($\ha_2$) and magnon mode ($\ha_1$).

The loop theorem applies to all types of optical cavities with multi-mode excitation, including the pure photonic systems \cite{Limo2017}, opto-mechanical systems, the opto-magnetic systems \cite{Auld1963,Chow1966,Weiner1972,Soykal2010,Cao2015,Yu2019,Huebl2013, Tabuchi2014, Zhang2014, Gor2014,Bai2015, Wang2018,Vahram2018, Harder2018, Bohi2019}, and even the classical coupled-oscillators.
There is also a rising interest in the coupling or entanglement of two or more macroscopic magnets under the assistance of a single cavity mode \cite{Xu2019,Borjans2019}. Our theory provides a simple yet systematic approach in describing and understanding all these systems.

{\it Conclusions.} In conclusion, we proposed a simple graphical loop theory to study the transmission spectrum of a multi-mode cavity. By drawing all the internal and external loops and ascertain their amplitudes, one can obtain the analytical expressions of the reflection and transmission spectrum without much calculation.

{\it Acknowledgements.}
This work was supported by the National Natural Science Foundation of China (Grants No. 61704071, No. 11722430, No. 11847202).
W.Y. is also supported by the China Postdoctoral Science Foundation under Grant No. 2018M641906.

\end{document}